\RequirePackage{fix-cm}
\documentclass[smallextended]{svjour3}       % onecolumn (second format)
\smartqed  % flush right qed marks, e.g. at end of proof
\usepackage{graphicx}
\begin{document}

\title{Quark model analysis of the Weinberg operator contribution to the nucleon EDM
%\thanks{Grants or other notes
%about the article that should go on the front page should be
%placed here. General acknowledgments should be placed at the end of the article.}
}
%\subtitle{Do you have a subtitle?\\ If so, write it here}

%\titlerunning{Short form of title}        % if too long for running head

\author{Nodoka Yamanaka         \and
        Emiko Hiyama %etc.
}

%\authorrunning{Short form of author list} % if too long for running head

\institute{N. Yamanaka \at
              Department of physics, Kennesaw State University, Kennesaw, GA 30144, USA \\
%              Tel.: +123-45-678910\\
%              Fax: +123-45-678910\\
              \email{nyamanak@kennesaw.edu}           %  \\
%             \emph{Present address:} of F. Author  %  if needed
           \and
           E. Hiyama \at
              Department of physics, Tohoku University, Sendai 980-8578, Japan
}

\date{Received: date / Accepted: date}
% The correct dates will be entered by the editor

\maketitle

\begin{abstract}
The Weinberg operator (chromo-electric dipole moment of gluon) is a CP violating quantity generated in many candidates of new physics beyond the standard model, and it contributes to observables such as the electric dipole moments (EDM) of the neutron or atoms which are currently measured in experiments.
In this proceedings contribution, we report on our result of the evaluation of the Weinberg operator contribution to the nucleon EDM in the nonrelativistic quark model using the Gaussian expansion method.
%Insert your abstract here. Include keywords, PACS and mathematical
%subject classification numbers as needed.
% \PACS{PACS code1 \and PACS code2 \and more}
% \subclass{MSC code1 \and MSC code2 \and more}
\end{abstract}

\section{Introduction}
\label{intro}

In the early era of the Universe, there were equal amounts of matter and antimatter, as well as particles or states which carry both baryon and lepton numbers.
When the temperature of the Universe crossed the energy of the latter states, the baryon and lepton number changing reactions stopped.
It is possible to generate the baryon number asymmetry at this timing if the theory is violating the C and CP symmetries, in addition to the violation of the baryon number and the departure from thermal equilibrium (Sakharov's criteria \cite{Sakharov:1967dj}).
This asymmetry is then converted to the baryon-to-photon ratio after the freezing of the pair-creations of matter and antimatter, which is nowadays 1:10$^{10}$.
It is however known that the standard model (SM) is not able to explain this large asymmetry due to the lack of CP violation \cite{Kobayashi:1973fv,Huet:1994jb}, and a new theory beyond it is absolutely necessary to solve this puzzle.

One of the most sensitive observables of CP violation beyond the SM is the electric dipole moment (EDM), which is measurable in many systems \cite{Yamanaka:2014mda,Yamanaka:2017mef,Chupp:2017rkp}.
It is defined as the permanent dipole polarization, and for the nucleon with three nonrelativistic quark, it is given as
\begin{eqnarray}
d_{N}
&=&
\sum_{i=1}^{3} \frac{e}{2} 
\langle \, \Psi_N \, |\, \tau_i^z \, {\cal R}_{iz} \, | \, \Psi_N \, \rangle
,
\label{eq:nucleonedmpolarization}
\end{eqnarray}
where $| \, \Psi_N \, \rangle$ is the wave function of the nucleon, and $\tau_i^z , {\cal R}_{iz}$ are the isospin and the coordinates of the $i$th quark, respectively.
The EDM may have a finite value only if there is a CP-odd component in $| \, \Psi_N \, \rangle$, which is induced by some CP violating interaction.
An important example is Weinberg's gluonic  dimension-6 operator \cite{Weinberg:1989dx,Bigi:1990kz,Bigi:1991rh}
\begin{eqnarray}
{\cal L}_w 
&=& 
\frac{1}{3!} w 
f^{abc} \epsilon^{\alpha \beta \gamma \delta} G^a_{\mu \alpha } G_{\beta \gamma}^b G_{\delta}^{\ \ \mu,c}
,
\label{eq:weinberg_operator}
\end{eqnarray}
which appears in many candidates of new physics beyond the SM, while it is very small in the SM \cite{Pospelov:1994uf,Yamaguchi:2020dsy}.
The EDM of the neutron was recently measured in experiment, with the upper limit \cite{Abel:2020gbr}
\begin{eqnarray}
|d_n| < 1.8 \times 10^{-26} e \, {\rm cm}
.
\end{eqnarray}
This is a very tight bound which may constrain the Weinberg operator coupling $w$ and the new physics behind it.
However, the relation between $w$ and $d_n$ has a large theoretical uncertainty due to the nonperturbative physics of QCD, and it is still difficult to quantify it in lattice QCD \cite{Dragos:2019oxn,Rizik:2020naq} or in perturbative QCD \cite{Hatta:2020ltd,Hatta:2020riw}.

\begin{figure*}
\includegraphics[width=0.75\textwidth]{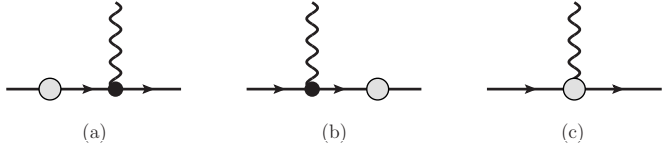}
\caption{Schematic picture of the nucleon EDM generated by the chiral rotation (a,b) and by the irreducible EDM (c).}
\label{fig:nucleon_EDM_Weinberg_operator}
\end{figure*}

In previous works, $d_n/w$ was evaluated using QCD sum rules \cite{Demir:2002gg,Haisch:2019bml}.
To be more precise, the CP-odd nucleon mass generated by the Weinberg operator of Eq.  (\ref{eq:weinberg_operator}) 
\begin{eqnarray}
{\cal L}
=
-m_{CP} \bar \psi_N i \gamma_5 \psi_N
,
\end{eqnarray}
has been calculated in QCD sum rules.
The nucleon EDM is then obtained by redefining the anomalous magnetic moment $\mu_N$ via chiral rotation, as \cite{Bigi:1990kz,Bigi:1991rh}
\begin{eqnarray}
d_N^{\rm (red)} (w) 
=
\frac{m_{CP}}{m_N} \mu_N
.
\end{eqnarray}
This is a one-particle reducible effect and it may be schematically drawn as Figs. \ref{fig:nucleon_EDM_Weinberg_operator} (a,b).
The most recent analysis yields \cite{Haisch:2019bml}
\begin{eqnarray}
d_N^{\rm (red)} (w) 
&\approx&
w \times 
\left\{
\begin{array}{rl}
 (25\pm 13) \, e \, {\rm MeV} & (N = n ) \cr
(-23\pm 12) \, e \, {\rm MeV} & (N = p ) \cr
\end{array}
\right.
.
\label{eq:weinbergop_red}
\end{eqnarray}
In previous works, the irreducible contribution [see Fig. \ref{fig:nucleon_EDM_Weinberg_operator} (c)], however, was neglected.
It is actually possible to calculate the irreducible EDM of the nucleon in the nonrelativistic quark model using the Gaussian expansion method \cite{Hiyama:2003cu}, as was done for the nuclear EDM  \cite{Yamanaka:2015qfa,Yamanaka:2015ncb,Yamanaka:2016itb,Yamanaka:2016fjj,Yamanaka:2016umw,Lee:2018flm,Yamanaka:2019vec}.
In this proceedings contribution, we report on the calculation of the irreducible contribution \cite{Yamanaka:2020kjo}.

\section{Quark model calculation of the nucleon EDM}
\label{sec:quarkmodel}

\subsection{Setup of the quark model}
\label{sec:setup}

In our calculation, we assume nonrelativistic quarks which interact through phenomenological potentials.
We analyze with three different (CP-even) interquark potentials (Bhaduri \cite{Bhaduri:1981pn}, AP1 and AL1 \cite{Semay:1994ht}).
The systematics of our analysis is estimated with the variation of the results obtained from them.
The potential of Bhaduri \cite{Bhaduri:1981pn} is given by
\begin{equation}
V_{qq , ij} (r)
=
-\frac{3}{4}
(t_a)_i \otimes (t_a)_j
\Biggl[
-\frac{\kappa}{r} 
+\lambda r 
+\Lambda
+\frac{\kappa}{m_Q^2} \frac{e^{- r /r_0}}{r r_0^2} \vec{\sigma}_i \vec{\sigma}_j
\Biggr]
,
\label{eq:bhaduri}
\end{equation}
where $r$ is the relative coordinate between the $i$th and $j$th quarks, $\vec{\sigma}$ and $t_a$ are the spin operator and the color generator respectively.
The other parameters are $m_Q = 337$ MeV, $\kappa = 0.52$, $\Lambda = - 0.9135$ GeV, $\lambda = 0.186 $ GeV$^2$, and $r_0 = 2.305$ GeV$^{-1}$ \cite{Bhaduri:1981pn}.
The AL1 and AP1 potentials are defined by \cite{Semay:1994ht}
\begin{equation}
V_{qq , ij} (r)
=
-\frac{3}{4}
(t_a)_i \otimes (t_a)_j
\Biggl[
-\frac{\kappa }{r} 
+\lambda r^p 
+\Lambda
+\frac{2 \pi \kappa'}{3 m_Q^2} \frac{e^{- r^2 /r_0^2}}{\pi^{3/2} r_0^3} \vec{\sigma}_i \vec{\sigma}_j
\Biggr]
,
\label{eq:semay}
\end{equation}
with $r_0 \equiv A / m_Q^B $.
For the parameters used, see Table \ref{tab:semay}.

\begin{table}
\caption{
Input parameters of the potentials AL1 and AP1 \cite{Semay:1994ht}.
The units of $m_Q$, $\lambda$, $\Lambda$, and $A$ are MeV, GeV$^{1+p}$, GeV, and GeV$^{B-1}$, respectively.
The others are dimensionless.
}
\label{tab:semay}
\begin{tabular}{l|cccccccc|}
\hline\noalign{\smallskip}
 & $p$ & $m_Q$ & $\kappa$ & $\kappa'$ & $\lambda$ & $\Lambda$ & $B$ & $A$ \\ 
\noalign{\smallskip}\hline\noalign{\smallskip}
AL1 & 1 & 315 & 0.5069 & 1.8609 & 0.1653 & -0.8321 & 0.2204 & 1.6553 \\
AP1 & 2/3 & 277 & 0.4242 & 1.8025 & 0.3898 & -1.1313 & 0.3263 & 1.5296 \\
\noalign{\smallskip}\hline
\end{tabular}
\end{table}

\begin{figure}
\includegraphics[width=0.25\textwidth]{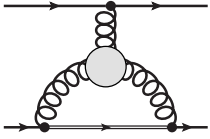}
\caption{One-loop level contribution of the Weinberg operator to the CP-odd interquark force.}
\label{fig:2qint}
\end{figure}

The CP-odd interquark force generated by the Weinberg operator (\ref{eq:weinberg_operator}) is derived by calculating the one-loop level diagram of Fig. \ref{fig:2qint} with the heavy quark approximation.
The gluon is also assumed to have a mass, $m_g =350$ MeV, obtained from Landau gauge lattice QCD \cite{Falcao:2020vyr}.
The CP-odd potential is then \cite{Yamanaka:2020kjo}
\begin{equation}
{\cal H}_{CPV ,ij}
=
-
\frac{N_c g_s \alpha_s m_g }{2 }
w
(\vec{\sigma}_i- \vec{\sigma}_j) 
\cdot
\vec{\nabla} \frac{e^{- m_g r }}{4 \pi r}
(t_a)_i \otimes (t_a)_j
,
\label{eq:cpvhamiltonian}
\end{equation}
having a close form to the CP-odd nuclear force with $\omega$ meson exchange \cite{Yamanaka:2015qfa,Yamanaka:2016umw,Froese:2021civ}.

\subsection{Gaussian expansion method}
\label{sec:GEM}

The nucleon is assumed to be a nonrelativistic three-body system, and its wave function is calculated by solving the Schr\"{o}dinger equation
\begin{eqnarray}
\Biggl[ 
\sum_i T_i+ \sum_{i \ne j} V_{qq , ij}
+{\cal H}_{CPV , ij} - E 
\Biggr] \, \Psi_{JM}(n,p)  = 0 
,
\label{eq:schr7}
\end{eqnarray}
where $T$ and $V$ are the kinetic term and the interquark potential introduced in Sec. \ref{sec:setup}.
For this purpose, we use the Gaussian expansion method \cite{Hiyama:2003cu} which is based on the variational principle.

\begin{figure*}
\includegraphics[width=0.75\textwidth]{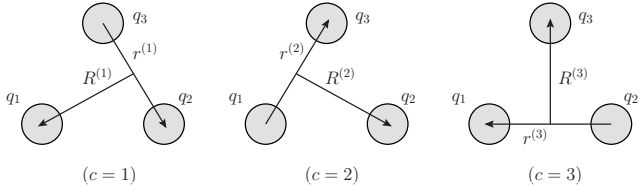}
\caption{Jacobi coordinates of the three-quark system.}
\label{fig:jacobi}
\end{figure*}

The nucleon wave function is given by the Jacobi coordinates depicted in Fig. \ref{fig:jacobi}, and it is explicitly written as
\begin{eqnarray}
&&\Psi_{JM}(n, p)
=
\sum_{c=1}^{3} \:
\sum_{nl, NL}
\sum_{\Sigma ,s}
\sum_{T} 
C^{(c)}_{nl,NL, \Sigma s, T}\: {\cal A} \Biggl[
\Bigl[  \eta^{(c)} ( T_c ) \otimes \eta'^{(c)} ({\scriptstyle \frac{1}{2}} ) \Bigr]_{I={\scriptstyle \frac{1}{2}} ,I_z}
\times 
\nonumber  \\
&&
\Bigl[
[ \phi^{(c)}_{nl}({\bf r}_c) \otimes \psi^{(c)}_{NL}({\bf R}_c)]_\Lambda \otimes \, \bigl[ \chi^{(c)} ( s_c ) \otimes \chi'^{(c)} ({\scriptstyle \frac{1}{2}} ) \bigr]_\Sigma \Bigr]_{J={\scriptstyle \frac{1}{2}}, M}
\hspace{-1em}
\times
| {\rm color\, singlet} \rangle
\Biggr]
,
\ \ \ \ \ 
\label{eq:he7lwf}
\end{eqnarray}
where ($\eta$,$\eta'$), ($\phi$,$\psi$), and ($\chi$,$\chi'$) are the nucleon isospin, radial, and spin functions, with the total antisymmetrizer $\cal{A}$.
We note that the color function is totally antisymmetric.
The radial functions $\phi_{nlm}({\bf r}) \equiv r^l \, e^{-(r/r_n)^2} Y_{lm}({\widehat {\bf r}})$ and $\psi_{NLM}({\bf R}) \equiv R^L \, e^{-(R/R_N)^2} Y_{LM}({\widehat {\bf R}})$ are expanded using the geometric progression $\rho_n = \rho_{\rm min} a^{n-1} \quad (n=1 - n_{\rm max} , \rho =r,R)$, up to angular momenta $l, L, \Lambda \leq 2$.
This three-body calculation exactly follows the same procedure as the nuclear EDM calculations \cite{Yamanaka:2015qfa,Yamanaka:2016umw,Yamanaka:2015ncb,Lee:2018flm,Yamanaka:2019vec}.
The final result for the irreducible nucleon EDM is obtained by calculating Eq. (\ref{eq:nucleonedmpolarization}) with the wave function obtained by solving Eq. (\ref{eq:schr7}).

\section{Result and summary}
\label{sec:result}

From our quark model calculation, we obtain the following result for the irreducible nucleon EDM [Fig. \ref{fig:nucleon_EDM_Weinberg_operator} (c)]
\begin{eqnarray}
d_N^{\rm (irr)} (w) 
&\approx&
\left\{
\begin{array}{rl}
-w \times (4-5) \, e \, {\rm MeV} & (N = n ) \cr
 w \times (4-5) \, e \, {\rm MeV} & (N = p ) \cr
\end{array}
\right.
,
\label{eq:weinbergop_result}
\end{eqnarray}
where the error bar is due to the choice of the interquark potential (see Sec. \ref{sec:setup}).
By combining with the QCD sum rules result (\ref{eq:weinbergop_red}) of Ref. \cite{Haisch:2019bml}, we obtain
\begin{eqnarray}
d_N (w) 
&=&
d_N^{\rm (red)} (w)
+ 
d_N^{\rm (irr)} (w) 
\nonumber\\
&=&
w\times
\left\{
\begin{array}{rl}
 ( 20 \pm 12)\, e \, {\rm MeV} & (N = n ) \cr
- ( 18 \pm 11)\, e \, {\rm MeV} & (N = p ) \cr
\end{array}
\right.
.
\label{eq:weinbergop_total}
\end{eqnarray}
We see that the EDM generated by the chiral rotation is dominant.
Since the latter is determined with an error bar of 50\% \cite{Haisch:2019bml}, the total nucleon EDM generated by the Weinberg operator is determined with an uncertainty of about 60\%, even if we assume an $O(100\%)$ error for our quark model calculation.
This is the first quantified result for the Weinberg operator.
We expect to also apply this analysis to the calculation of the CP-odd nuclear force which is an important input in the analysis of the atomic and nuclear EDMs \cite{Yamanaka:2017mef,Chupp:2017rkp,Yamanaka:2016umw}.

\end{document}